\documentstyle{article}
\input{psfig.tex}
\title{Cosmic Inflation\footnote{
Lectures presented at the NATO Advanced Studies Institute ``Structure
Formation in the Universe'', Cambridge 1999.  To be published by
Kluwer, R. Crittenden and N. Turok Eds.
}
}
\author{Andreas Albrecht\\ U. C. Davis Department of Physics\\
           One Shields Avenue, Davis, CA, 95616}
\begin{document}
\maketitle

\centerline{\bf Abstract}
I review the current status of the theory of Cosmic Inflation.  My
discussion covers the motivation and implementation of the idea, as
well as an analysis of recent successes and open questions.   There is
a special discussion of the physics of ``cosmic coherence'' in the
primordial perturbations.  The issue of coherence is at the heart of
much of the success inflation theory has achieved at the hands of
the new microwave background data. While much of this review should be
useful to anyone seeking to update  their knowledge of inflation
theory, I have also made a point of including basic material directed at
students just starting research.  

\bigskip
\bigskip
\bigskip
\bigskip
\bigskip
\bigskip
\bigskip
\bigskip
\bigskip
\bigskip
\bigskip
\bigskip
\bigskip
\bigskip
\bigskip
\bigskip
\bigskip
\bigskip
\bigskip
\bigskip
\bigskip
\pagebreak
\tableofcontents

\section{Introduction}

It has been almost 20 years since Guth's original paper fueled considerable
excitement in the cosmology community \cite{Guth}.  It was through this
paper that many cosmologists saw the first glimmer of hope that
certain deep mysteries of the Universe could actually be resolved
through the idea of Cosmic Inflation.  Today inflation theory has come
a long way.  While at first the driving forces were deep theoretical
questions, now the most exciting developments have come at the hands
of new cosmological observations.  In the intervening period we have
learned a lot about the predictions inflation makes for the state of
the Universe today, and new technology is allowing these predictions
to be tested to an ever growing degree of precision.  The continual
confrontation with new observational data has produced a string of
striking successes for inflation which have increased our confidence
that we are on the right track. Even so, there are key unanswered
questions about the foundations of inflation which become ever more
compelling in the face of growing observational successes. 

\section{Motivating Inflation: The Cosmological Features/Problems}
\label{MITCFP}
It turns out that most of the cosmological ``problems'' that are usually
introduced as a motivation for inflation are actually only
``problems'' if you take a very special perspective.  I will take
extra care to be clear about this by first presenting these issues simply
as ``features'' of the standard Big Bang, and then discussing 
circumstances under which these features can be regarded as ``problems''.  

\subsection{The Standard Big Bang}
To start with, I will introduce the basic tools and ideas of the
standard ``Big Bang'' (SBB) cosmology.  Students needing further
background should consult a text such as Kolb and Turner  \cite{KT}. An
extensive discussion of inflation can 
be found in  \cite{LindeBook}. The SBB treats a nearly perfectly
homogeneous and isotropic universe, which gives a good fit to present
observations. The single dynamical parameter describing the broad
features of the SBB is the ``scale factor'' $a$, which obeys the
``Friedmann equation''
\begin{equation}
\left({\dot a \over a}\right)^2 \equiv H^2 ={8\pi\over 3}\rho-{ k
\over a^{2}} 
\label{FE}
\end{equation}
in units where $M_P = \hbar = c = 1$, $\rho$ is the energy density
and $k$ is the curvature.
The Friedmann equation can be solved for $a(t)$ once $\rho(a)$ is
determined.  This can be done using local energy conservation, which
for the SBB cosmology reads
\begin{equation}
{d \over da} \left( \rho a^3\right) = -3 {p }{d \over da} a^3.
\label{econs}
\end{equation}
Here (in the comoving frame) the stress energy tensor of the matter is 
given by
\begin{equation}
T^{\mu}_{\nu} = Diag(\rho, p, p, p).
\label{Tflu}
\end{equation}
In the SBB, the Universe is first dominated by relativistic matter
(``radiation dominated'') with $w=1/3$ which gives $\rho \propto
a^{-4}$ and $a \propto t^{1/2}$. Later the Universe is dominated by 
non-relativistic matter (``matter dominated'') with $w=0$ which gives $\rho
\propto a^{-3}$ and $a \propto t^{2/3}$. 

The scale factor measures the overall expansion of the Universe
(it doubles in size as the separation between distant objects
doubles). Current data suggests an additional ``Cosmological
Constant'' term $\equiv \Lambda/3$ might be present on the right hand
side of the Friedmann equation with a size similar to the other terms.
However with the $\rho$ and $k$ terms evolving as
negative powers of $a$ these terms completely dominate over $\Lambda$
at the earlier epochs we are discussing here.  We
set $\Lambda=0$ for the rest of this article.

\subsection{The Flatness Feature}
The ``critical density'', $\rho_c$, is defined by
\begin{equation}
H^2 = \left({\dot a \over a}\right)^2 ={8\pi\over 3}\rho_c.
\end{equation}
A universe with $k=0$ has $\rho = \rho_c$ and is said to be ``flat''.
It is useful to define the dimensionless density parameter 
\begin{equation}
\Omega \equiv {\rho \over \rho_c}.
\label{kofomega}  
\end{equation}
If $\Omega$ is close to unity the $\rho$ term dominates in the
Friedmann equation and the Universe is nearly flat.  If $\Omega$
deviates significantly from unity the  $k$ term (the ``curvature'') is
dominant. 

\begin{figure}
\centerline{\psfig{file=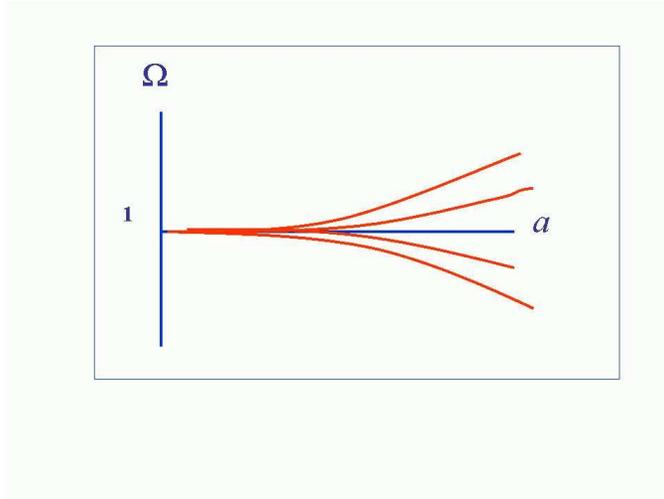,width=3.5in}}
\caption{In the SBB $\Omega(a)$ tends to evolve away from unity as the
Universe expands} 
\label{flatness}
\end{figure}

The Flatness feature stems from the fact that $\Omega = 1$ is an
unstable point in the evolution of the Universe.  Because $\rho
\propto a^{-3}$ or $a^{-4}$ throughout the history of the Universe, the
$\rho$ term in the Friedmann equation falls away much more quickly than 
the $k/a^2$ term as the Universe expands, and the $k/a^2$ comes to
dominate.  This behavior is
illustrated in Fig. \ref{flatness}.

Despite the strong tendency for the equations to drive the Universe
away from critical density, the value of $\Omega$ today is remarkably
close to unity even after 15 Billion years of evolution.  Today the value of 
$\Omega$ is within an order of magnitude of unity, and that
means that at early times $\rho$ must have taken values that were set
extremely closely to $\rho_c$.  For example at the epoch of Grand
Unified Theories (GUTs) ($T \approx  10^{16} GeV$), $\rho$ has to
equal $\rho_c$ 
to around 55 decimal places.  This is simply an important property $\rho$ must
have for the SBB to fit the current observations, and at this point we 
merely take it as a feature (the ``Flatness Feature'') of the SBB.

\subsection{The Homogeneity Feature}
The Universe today is strikingly homogeneous on large scales.
Observations of the Cosmic Microwave Background (CMB) anisotropies
show that the Universe was even more inhomogeneous in the distant past.
Because the processes of gravitational collapse causes inhomogeneities
to {\em grow} with time, an even greater degree of homogeneity must
have been present at earlier epochs. 

As with the flatness feature, the equations are driving the Universe
into a state (very inhomogeneous) that is very different from the way
it is observed today. This difference is present despite the fact
that the equations have had a very long time to act. To arrange this
situation, very special (highly homogeneous) initial conditions must
be chosen, so that excessive inhomogeneities are not 
produced.

One might note that at early times the Universe was a hot
relativistic plasma in local equilibrium and ask:  Why can't
equilibration processes smooth out any fluctuations and produce the
necessary homogeneity?  This idea does not work because equilibration
and equilibrium are only possible on scales smaller than the ``Jeans
Length'' ($R_J$). On larger scales gravitational collapse wins over
pressure and drives the system into an inhomogeneous non-equilibrium
state.  

At early epochs the region of the Universe we observe today
occupied many Jeans volumes.  For example, at the epoch of Grand
Unification the observed Universe contained around $10^{80}$ Jeans
volumes (each of which would correspond to a scale of only several
$cm$ today). So the tendency toward gravitational collapse has had a
chance to have a great impact on the Universe over the course of its
history. Only through extremely precisely determined initial
conditions can the amplitudes of the resulting inhomogeneities be
brought down to acceptable levels.
At the GUT epoch, for example, initial density contrasts
\begin{equation}
\delta(x) \equiv (\rho(x) - \bar\rho)/\bar \rho
\label{deltadef}
\end{equation}
that are 
non-zero only in the 20th, 40th or even later decimal places,
depending on the scale, are 
required to arrange the correct degree of homogeneity.  
So here we have another special feature, the ``Homogeneity Feature''
which is required to make the SBB consistent with the Universe that we 
observe. 

\subsection{The Horizon Feature}
One can take any physical system and choose 
an ``initial time'' $t_i$.  The state of the system at any later time
is affected both by the state at $t_i$ (the initial conditions) and by 
the subsequent evolution.  At any finite time after $t_i$ there are causal 
limits on how large a scale can be affected by the subsequent
evolution (limited ultimately by the speed of light, but really by
the actual propagation speeds in that particularly system, which can
be much slower).  So there are always sufficiently large scales which
have not been affected by the subsequent evolution, and on which the
state of the system is simply a reflection of the initial conditions.

This same concept can equally well be applied to the SBB.  However,
there is one very interesting difference:  The SBB starts with an
initial singularity where $a=0$ and $\rho= 
\infty $ at a finite time in the past.  This is a serious enough
singularity that the equations cannot propagate the Universe through
it.  In the SBB the Universe must simply start at this initial
singularity with chosen initial conditions.  So while in a laboratory
situation the definition of $t_i$ might be arbitrary in many cases,
there is an absolutely defined $t_i$ in the SBB.  

\begin{figure}
\centerline{\psfig{file=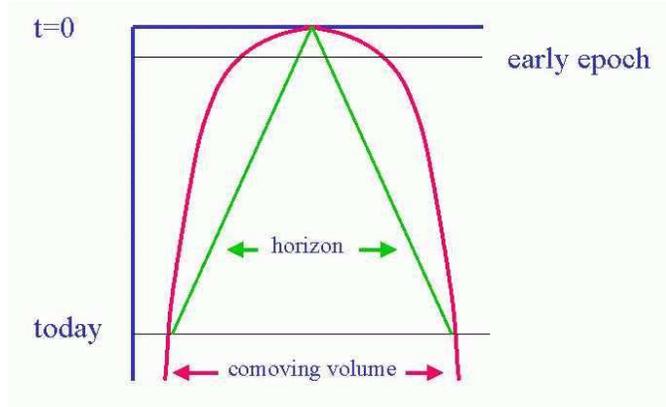,width=3.5in}}
\caption{The region of the Universe we see today was composed of many
causally disconnected regions in the past.} 
\label{horizon}
\end{figure}

As with any system, the causal ``horizon'' of the Universe grows with
time.  Today, the region with which we are just coming into causal
contact (by observing distant points in the Universe) is one ``causal
radius'' in size, which means objects we see in opposite directions
are two causal radii apart and have not yet come into causal
contact. One can calculate the number of causal regions that filled
the currently observed Universe at other times.  In the early Universe 
the size of a causally connected region was roughly the Jeans length,
and so at the Grand Unification epoch there were around $10^{80}$
causally disconnected regions in the volume that would evolve into the 
part of the Universe we currently can observe.  This is the ``Horizon
Feature'' of the SBB, and is depicted in Fig. \ref{horizon}

\subsection{When the Features become Problems}
I have discussed three features of the SBB that are usually presented
as ``problems''.  What can make a feature a problem?  The Flatness and
Homogeneity features describe the need to give the SBB very special
initial conditions in order to match current data.  Who cares?  If you 
look at the typical laboratory comparison between theory and
experiment, success is usually measured by whether or not theoretical
 equations of motion correctly describe the evolution of the
system.  The choice of initial conditions is usually made only in
order to facilitate the comparison.  By these standards, the SBB does
not have any problems.  The equations, with suitably chosen initial
conditions, do a perfectly good job of describing the evolution of the 
Universe.  

However, there is something extremely strange about the initial
conditions required by the SBB that makes us unable to simply accept
the Flatness and Homogeneity features:  The required initial
conditions place the Universe very far away from where its equations
of motion wish to take it.  So much so that even today, 15 Billion
years later, the Universe still has features (flatness and homogeneity) 
that the equations are trying to destroy.

I think a simple analogy is useful:  Imagine that you come into my
office and see a pencil balanced on its point.  You notice it is still 
there the next day, the next week, and even a year later.  You finally 
ask me ``what is going on with the pencil?'' to which I reply: ``yes,
that's pretty interesting, but I don't have to explain it because it
was that way when I moved into my office''.  You would find such a
response completely unreasonable.  On the other hand, if the pencil
was simply lying on my desk, you probably would not care at all how it 
got there.

The fact is that, like the fictitious balanced pencil, the initial conditions
of the real Universe are so amazing that we
cannot tear ourselves away from trying to understand what created
them.  It is at this point that the Flatness and Homogeneity features
become ``problems''.

It is also at this point that the Horizon feature becomes a problem.
The hope of a cosmologist who is trying to explain the special initial 
conditions is that some new physical processes can be identified which 
can ``set up'' these initial conditions.  A natural place to hope for
this new physics to occur is the Grand Unification (GUT) scale, where
particle physics already indicates that something new will happen.
However, once one starts thinking that way, one faces the fact that the
Universe is compose of $10^{80}$ causally disconnected regions at the
GUT epoch, and is seems impossible for physical processes to extend
across these many disconnected regions and set up the special
homogeneous initial conditions. 

\subsection{The Monopole Problem}
There is no way the Monopole problem can be regarded as merely a
feature.  In 1979 Preskill  \cite{Preskill79} showed how quite
generally a significant number of magnetic monopoles would be produced 
during the cosmic phase transition required by Grand Unification.
Being non-relativistic ($\rho \propto a^{-3}$), they would 
rapidly come to dominate over the ordinary matter (which was then relativistic,
$\rho \propto a^{-4}$).  Thus it appeared that a Universe that is
nearly 100\%  magnetic monopoles today was a robust prediction from GUT's, and thus 
the idea of Grand Unification was easily ruled out.

\section{The Machinery of Inflation and what it can do}

Over the years numerous inflationary ``scenarios'' have been
proposed.  My goal here is to introduce the basic ideas that underlie
essentially all the scenarios.

\subsection{The potential dominated state}
Consider adding an additional scalar matter field $\phi$ (the ``inflaton'') to
the cosmological model.  In general the stress-energy tensor of this
field will depend on the space and time derivatives of the field as
well as the potential $V(\phi)$.  In the limit where the potential
term dominates over the derivative terms the stress-energy tensor of
$\phi$ takes the form
\begin{equation}
T^{\mu}_{\nu} = Diag(V,-V, -V, -V)
\label{TV}
\end{equation}
which leads (by comparison with Eqn. \ref{Tflu}) to $\rho = -p = V$.

Solving Eqn. \ref{econs} gives $\rho(a) = V$.  Because we are assuming 
$\dot \phi$ is negligible we also have $V = {\rm const.}$.  Under
these circumstances $H = {\rm const.}$ and the Friedmann equation
gives exponential expansion:
\begin{equation}
a(t) \propto e^{Ht},
\end{equation}
very different from the power law behavior in the SBB.

\subsection{A simple inflationary cosmology: Solving the problems}
In the early days the inflationary cosmology was seen as
a small modification of the SBB:  One simply supposed that there was
a period (typically around the GUT epoch) where the Universe
temporarily entered a potential dominated state and ``inflated''
exponentially for a period of time. Things were arranged so that
afterwards all the energy coupled out of $V(\phi)$ and into ordinary
matter (``reheating'') and the SBB continued normally.

\begin{figure}
\centerline{\psfig{file=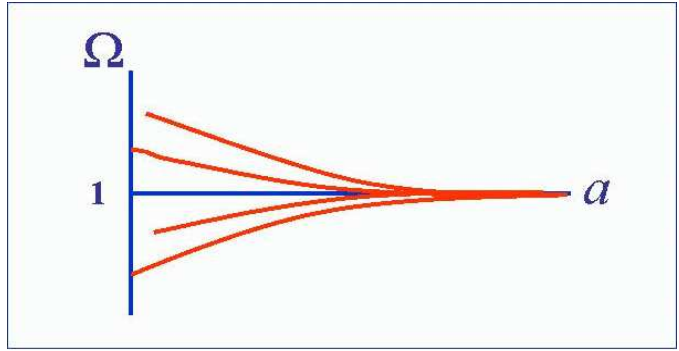,width=3.5in}}
\caption{During inflation $\Omega = 1$ is an attractor} 
\label{flat-inf}
\end{figure}

This modification on the SBB has a profound effect on the cosmological 
problems.  I will address each in turn:

{\bf Flatness:}  The flatness problem was phrased in terms of the
tendency for the $k$ term to dominate the $\rho$ term in the Friedmann 
equation (Eqn. \ref{FE}).  During inflation with $\rho = {\rm const.}$ 
the reverse is true and the $k$ term becomes negligible, driving the
Universe {\em toward} critical density and $\Omega=1$.  This process
is illustrated in Fig. \ref{flat-inf}.

\begin{figure}
\centerline{\psfig{file=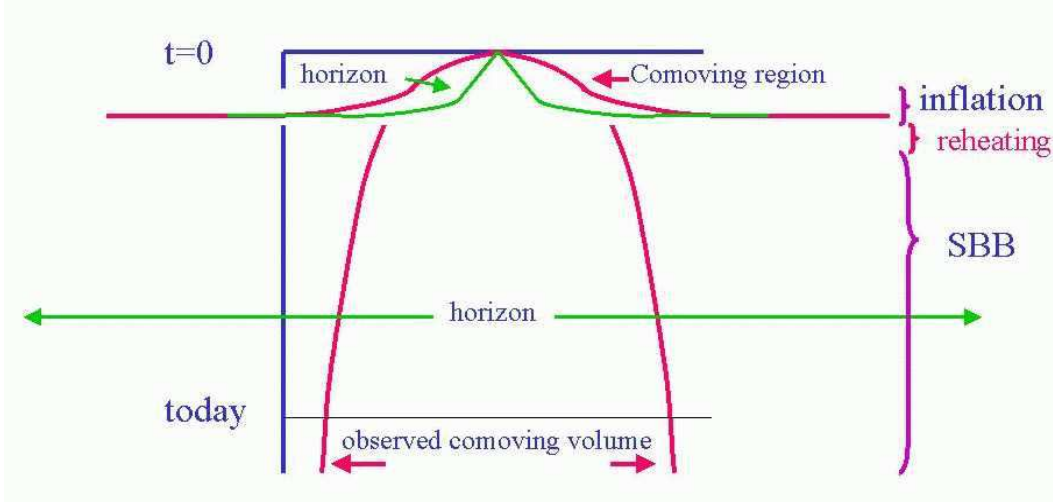,width=5.5in}}
\caption{During inflation one horizon volume is exponentially
expanded.  Today, the observe region can occupy a small fraction of a
causal volume.} 
\label{horiz-inf}
\end{figure}

{\bf Horizon:}  A period of inflation radically changes the causality
structure of the Universe. During inflation, each causally connected region is
expanded exponentially.  A suitable amount of inflation allows
the entire observed Universe to come from a region that was causally
connected before inflation, as depicted in
Fig. \ref{horiz-inf}. Typically one needs the Universe to 
expand by a factor of around $e^{60}$ to solve the Horizon and
Flatness problems with GUT epoch inflation.  

{\bf Homogeneity:}  Of course ``solving'' the Horizon problem does not
guarantee a successful picture.  Bringing the entire observable Universe 
inside one causally connected domain in the distant past can give one
{\em hope} that the initial conditions of the SBB can be explained by physical
processes.  Still, one must determine exactly what the relevant physical
processes manage to accomplish.  While the result of inflation is
clearcut in terms of the flatness, it is much less so in terms of the
homogeneity.   The good news is that a given inflation
model can actually make {\em predictions} for the spectrum of
inhomogeneities that are present at the end of inflation.  However,
there is nothing intrinsic to inflation that predicts that these
inhomogeneities are small.  None the less, the mechanics of inflation
can be further adjusted to give inhomogeneities of the right
amplitude.  Interestingly, inflation has a lot to say about other
aspects of the early inhomogeneities, and these other aspects are 
testing out remarkably well in the face of new data.  There will be
more discussion of these matters in Sections \ref{Perturbations} and
\ref{CosmicCoherence}.  

{\bf Monopoles:} The original Monopole problem occurs because GUTs
produce magnetic monopoles at sufficiently high
temperatures.  In the SBB these monopoles ``freeze out'' in
such high numbers as the Universe cools that they rapidly
dominate over other matter (which is relativistic at that time).
Inflation can get around this problem if the reheating after inflation 
does not reach temperatures high enough to produce the monopoles.
During inflation, all the other matter (including any monopoles that
may be present) is diluted to completely negligible densities and the
overall density completely is dominated by $V(\phi)$.  The ordinary matter of
the SBB is created by the reheating process at the end of inflation.
The key difference is that in a simple SBB the matter we see has in
the past existed at all temperatures, right up to infinity at the
initial singularity.  With the introduction of an inflationary epoch,
the matter around us has only existed up to a finite maximum temperature
in the past (the reheating temperature). If this temperature is on the 
low side of the GUT temperature, monopoles will not be produced and
the Monopole Problem is evaded.

\subsection{Mechanisms for producing inflation}
Scalar fields play a key role in the standard mechanism of spontaneous 
symmetry breaking, which is widely regarded in particle physics as the 
fundamental origin of all particle masses.  Essentially all attempts
to establish a deeper picture of fundamental particles have introduced 
additional particles to the ones observed (typically to round out
representations of larger symmetry groups used to unify the
fundamental forces).  Conflicts with observations are then typically
avoided by giving the extra particles sufficiently large masses that
they could not be produced in existing accelerators.  Of course, this
brings scalar fields into play.

The upshot is that additional scalar fields abound, at least in the
imaginations of particle theorists, and if anything the problem for
cosmologists has been that there are too many different models.  It is 
difficult to put forward any one of them as the most compelling.  This 
situation has caused the world of cosmology to regard the ``inflaton'' 
in a phenomenological way, simply investigating the behaviors of
different inflaton potentials, and leaving the question of foundations 
to a time when the particle physics situation becomes clearer.

The challenge then is to account for at least $60$ e-folding of
inflation. Looking at the GUT scale (characterized by $M_G \approx M_P/1,000$) we can
estimate $H = \sqrt{(8 \pi/3) M_P^{-2} M_G^4}$, and a characteristic GUT 
timescale would be $t_g \approx M_G^{-1}$.  So one could estimate a
``natural'' number of e-foldings for GUT-scale inflation to be
$t_g\times H
\approx \sqrt{8 \pi/3}(M_G/M_P)$ where I have explicitly shown the
Planck Mass for clarity.  The upshot is that the ``natural'' number of 
e-foldings is a small fraction of unity, and there is the question of
how to get sufficient inflation.

\begin{figure}
\centerline{\psfig{file=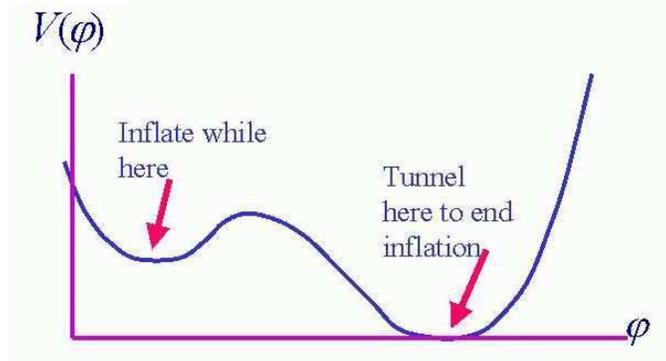,width=3.5in}}
\caption{In the original version of inflation proposed by Guth, the
potential dominated state was achieved in a local minimum of the
inflaton potential. Long periods of inflation were easily produced
because the timescale for inflation was set by a tunneling
process. However, these models had a ``graceful exit problem'' because 
the tunneling process produced a very inhomogeneous energy distribution.}
\label{oldinf}
\end{figure}

The ``sufficient inflation'' issue was easily addressed in the original Guth
paper \cite{Guth} because the end of inflation was brought on by an
exponentially long tunneling process with a timescale of the form $ t
\approx M_g^{-1} exp{(B)}$ where the dimensionless quantity $B$ need take
on only a modestly large value to provide sufficient inflation.  In
the Guth picture, the inflaton was trapped inside a classically stable
local minimum of the inflation potential, and only quantum tunneling
processes could end inflation.  This type of potential is depicted in
Fig. \ref{oldinf}.
The original picture proposed by Guth had serious problems however,
because the tunneling operated via a process of bubble nucleation.  This
process was analyzed carefully by Guth and Weinberg
 \cite{GuthWeinberg} and it was  
show that reheating was problematic in these models: The bubbles
formed with all the energy in their walls, and bubble collisions could
not occur sufficiently rapidly to dissipate the energy in a more
homogeneous way.  The original model of inflation had a
``graceful exit'' problem that made its prediction completely
incompatible with observations.

\begin{figure}
\centerline{\psfig{file=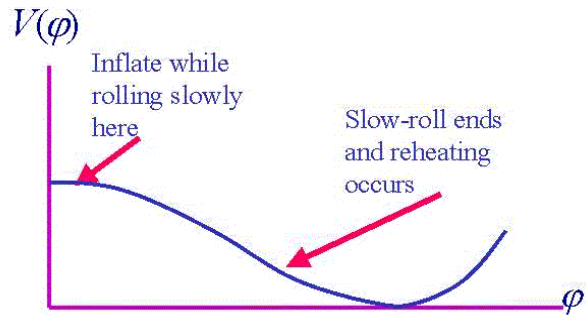,width=3.5in}}
\caption{In ``new'' inflation a classical instability ends inflation.
This allows for a graceful exit, but makes it more of a challenge to
get sufficiently long inflation times.} 
\label{newinf}
\end{figure}

The graceful exit problem was resolved by the idea of ``slow roll'' inflation
 \cite{Linde82,AlbrechtSteinhardt}.  In the slow roll picture inflation is {\em
classically} unstable, as depicted in Fig. \ref{newinf}.  
Slow roll
models are much easer to reheat and match onto the SBB.  However,
because the timescale of inflation is set by a classical instability,
there is no exponential working to your advantage. Fine tuning of
potential parameters is generally required to produce sufficient
inflation in slow roll models.  Essentially all current models of
inflation use the slow roll mechanism.  Interestingly, the constraints 
on the inflaton potential which are required to give the
inhomogeneities a reasonable amplitude tend to exceed the constraints
required to produce {\em sufficient} inflation.  Thus the requirement
of achieving sufficient inflation ends up in practice not providing
any additional constraint on model building. 

I have outlined the tunneling and slow roll pictures.  While the slow
roll remains the fundamental tool in all models, there are some
variations that incorporate aspects of both.  In some ``open
inflation'' models  \cite{BucherGoldhaberTurok} there is inflation with a
tunneling instability {\em and} a brief period of further inflation
inside the bubble after tunneling.  This arrangement can produce
values of $\Omega$ which are not necessarily close to unity. Also,
inflation has been studied in a multi-dimensional inflaton space where
there are some barriers and some slow roll directions.

\subsection{The start of inflation}
\label{tsoi}
In the original basic pictures of inflation, where a period of
inflation was inserted into a SBB model, various arguments were made
about why the Universe {\em had} to enter an inflationary phase based 
on symmetry restoration at high
temperatures \cite{Guth,Linde82,AlbrechtSteinhardt}. 
However, starting with a SBB {\em before} inflation assumes a great
deal about initial conditions (which after all we are trying to
explain).

Currently cosmologists that think about these things take one of two
perspectives. One group tries to treat
``the most random possible'' initial conditions \cite{LindeMezhlumian95}, and discuss how
inflation can emerge from such a state of matter. The other group
tries to construct the ``wavefunction of the Universe'' based on deep
principles \cite{HartleHawking}.  In some cases the wavefunction of the Universe indicates
that inflation is the most likely starting point for regions that take
on SBB-like behavior.  I have a personal prejudice against
``principles'' of initial conditions, because I do not see how that
gets your further than simply postulating the initial conditions of
the SBB.  

Coming from the random initial conditions point of view, a popular
argument is that even if the probability of starting inflation is
extremely small in the primordial chaos, once inflation starts it
takes over the Universe. When measured in terms of volume, the
exponential expansion really does appear to take over.  In fact, even
though there is a classical instability, the small quantum probability 
of remaining on the flat inflationary part of the potential is
amplified by the exponential inflationary growth, and in many models
most of the Universe inflates forever (so-called ``eternal
inflation'').  So perhaps as long as inflation is possible, one does
not even have to think too hard about what went before.

All of these issues are plagued by key unresolved questions (for
example to do with putting a measure on the large spaces one is
contemplating).  I will discuss some of the open questions associated
with these issues in Section \ref{tbpaoq}.

\subsection{Perturbations}
\label{Perturbations}
The technology for
calculating the production of
inhomogeneities during inflation is by now very well understood (For
an excellent review see  \cite{LiddleLyth}).  The
natural variables to follow are the Fourier
transformed density contrasts $\delta(k)$ in {\em comoving
coordinates} ($ k_{co} = a k_{phys}$).  A given mode $\delta(k)$
behaves very differently depending on its relationship to the Hubble
radius $R_H \equiv 1/H$.  During inflation $H = {\rm \it const.}$ while
comoving scales are growing exponentially.  The modes of interest
start inside $R_H$ ($k \gg R_H^{-1}$) and evolve outside $R_H$ ($k <
R_H^{-1}$) during inflation.  After inflation, during the SBB, $R_H
\propto t$ and $a \propto t^{1/2}$ or $t^{2/3}$ so $R_H$ catches up
with comoving scales and modes are said to ``fall inside'' the Hubble
radius.

It is important to note that $R_H$ is often called the
``horizon''.  This is because in the SBB $R_H$ is roughly equal to the 
distance light has traveled since the big bang.  Inflation changes all 
that, but ``Hubble radius'' and ``horizon'' are still often (confusingly)
used interchangeably.

Modes that are relevant to observed structures in the Universe started 
out on extremely small scales before inflation.  By comparison with
what we see in nature today, the only excitations one expects in these
modes are zero-point fluctuations in the quantum fields. Even if they
start out in an excited state, one might expect that on length-scales much
smaller (or energies-scales much higher) than the scale ($\rho^{1/4}$)
set by the 
energy density the modes would rapidly ``equilibrate'' to the ground
state.  These zero-point fluctuations are expanded to cosmic scales
during inflation, and form the ``initial conditions'' that are used to
calculate the perturbations today.  Because the perturbations on
cosmologically relevant scales must have been very small over most of
the history of the Universe, linear perturbation theory in GR may be
used.  Thus the 
question of perturbations from inflation can be treated by a tractable 
calculation using well-defined initial conditions.

One can get a feeling for some key features by examining this
result which holds for most inflationary models:
\begin{equation}
\delta_H = {32 \over 75} {16 \pi \over M_p^6 }{V_*^{3} \over
(V_*^{\prime})^{2}}.
\label{deltah}
%equations 5.7 and 5.21 from L&L
\end{equation}
Here $\delta_H^2$ is roughly the mean squared value of $\delta$ when
the mode falls below $R_H$ during the SBB. The ``$*$'' means evaluate
the quantity when the mode in question goes outside $R_H$ during
inflation, and $V$ is the inflaton potential.  Since typically the
inflaton is barely moving during inflation, $\delta_H^2$ is nearly
the same for all modes, resulting in a spectrum of perturbations that
is called ``nearly scale invariant''.  The cosmological data 
require  $\delta_H \approx 10^{-5}$. 

In typical models 
the ``slow roll'' condition, 
\begin{equation}
V'M_P/V \ll 1
\end{equation}
is obeyed, and inflation 
ends when $ V'M_P/V \approx 1$.  If $V$ is dominated by a single power 
of $\phi$ during the relevant period  $ V'M_P/V \approx 1$ means $\phi
\approx M_P$ at the end of inflation. Since $\phi$ typically does not
vary much during inflation $\phi 
\approx M_P$ would apply throughout. With these values of $\phi$, and
assuming simple power law forms for $V$, one might guess that the
right side of Eqn. \ref{deltah} is around unity (as opposed to
$10^{-5}$!) Clearly some subtlety is involved in constructing a
successful model. 

I have noticed people sometimes make rough dimensional arguments by
inserting $V \approx M_G^4$ and 
$V^{\prime} \approx M_G^3$ into Eqn. \ref{deltah} and conclude
$\delta_H \approx (M_G/M_P)^3$.  This argument suggests
that $\delta_H$ is naturally small for GUT inflation, and that the
adjustments required to get $\delta_H \approx 10^{-5}$ could be
modest. However that argument neglects the fact that the slow roll 
condition must be met. When this constraint is factored in, the
challenge of achieving the right amplitude is much greater than the
simplest dimensional argument would indicate.

\subsection{Further remarks}
Inflation provides the first known mechanism for producing the flat,
slightly perturbed initial conditions of the SBB.  Despite the fact
that there are many deep questions about inflation that have yet to be 
answered, it is still tremendous progress to know that nature has a
way around the cosmological problems posed in Section \ref{MITCFP}.
Whether nature has chosen this route is another question.
Fortunately, inflation offers us a number of observational signatures
that can help give an answer.  These signatures are the subject of
Sections \ref{CosmicCoherence} and \ref{Predictions}.

\section{Cosmic Coherence}
\label{CosmicCoherence}
In a typical inflation model the inflaton is completely out of the
picture by the end of reheating and one is left with a SBB model set up
with particular initial conditions. In realistic models the
inhomogeneities are small at early times and can be treated in
perturbation theory.  Linear perturbation theory in the SBB has
special properties that lead to a certain type of phase coherence in
the perturbations. This coherence can be observed, especially in the
microwave background anisotropies. Models where the perturbations are
subject to only linear evolution at early times are call ``passive'' models

\subsection{Key ingredients}
At sufficiently early times the Universe is hot and dense enough that
photons interact frequently and are thus tightly coupled to the
Baryons.  Fourier modes of the density contrast $\delta_{\gamma}(k)$ in the 
resulting ``photon-baryon fluid'' have the following 
properties:  On scales above the Jeans length the fluid experiences
gravitational collapse.  This is manifested by the presence of one
growing and one decaying solution for $\delta$.  One can think of the
growing solution corresponding to gravitational collapse and the
decaying solution corresponding to the evolution of an expanding
overdense region which sees its expansion slowed by gravity.  During
the radiation era Jeans length $R_J = R_H/\sqrt{3}$ and modes ``fall inside''
$R_J$ just as they fall inside $R_H$. For wavelengths smaller than
$R_J$ the pressure counteracts gravity and instead of collapse the
perturbations undergo oscillations, in the form of pressure (or
``acoustic'') waves. The experience of a given mode that starts outside
$R_J$ is to first experience gravitational collapse and then
oscillatory behavior.

There is a later stage when the photons and Baryons decouple (which we 
can think of here in the instantaneous approximation).  After
decoupling the photons ``free stream'', interacting only
gravitationally right up to the present day.  The process of first
undergoing collapse, followed by oscillation, is what creates the phase 
coherence.  I will illustrate this mechanism first with a toy model.

\subsection{A toy model}
\begin{figure}
\centerline{\psfig{file=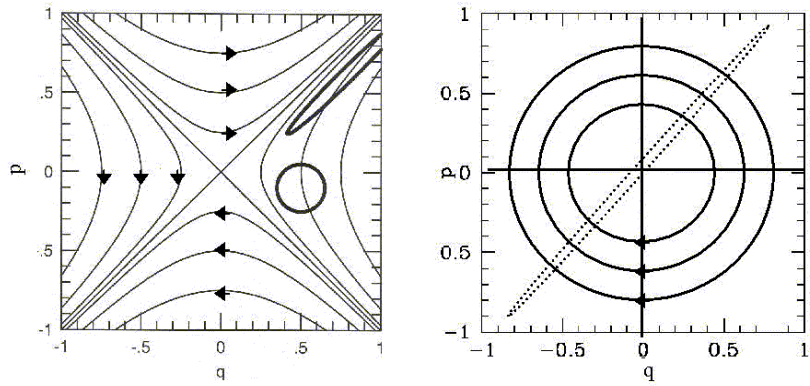,width=4.5in}}
\caption{The phase space trajectories for an upside-down harmonic
oscillator  are depicted in the left panel. Any region of phase 
space will be squeezed along the diagonal line as
the system evolves (i.e. the circle gets squeezed into the ellipse) .
For a right-side-up harmonic oscillator paths in the phase space are
circles, and angular position on the circle gives the phase of
oscillation.  Perturbations in the early universe exhibit first
squeezing and then oscillatory behavior, and any initial phase space
region will emerge into the oscillatory epoch in a form something like 
the dotted ``cigar'' due to the earlier squeezing. In this way the
early period of squeezing fixes the phase of oscillation.} 
\label{squeeze}
\end{figure}

During the ``outside $R_J$'' regime, there is one growing and one
decaying solution.  The simplest system which has this qualitative
behavior is the upside-down harmonic oscillator which obeys:
\begin{equation}
\ddot q = q.
\end{equation}
The phase space trajectories are show on the left panel in
Fig. \ref{squeeze}. The system is unstable against the runaway (or growing)
solution where $|q|$ and $|p|$ get arbitrarily large (and $p$ and $q$ have the
same sign). This behavior ``squeezes'' any initial region in phase
space toward the diagonal line with unit slope.  The squeezing effect
is illustrated by the circle which evolves, after a period of time,
into the ellipse in Fig. \ref{squeeze}.

The simplest system showing oscillatory behavior is the normal
harmonic oscillator obeying
\begin{equation}
\ddot q = -q.
\end{equation}
This phase space trajectories for this system are circles, as shown in 
the right panel of Fig. \ref{squeeze}. The angular position around the 
circle corresponds to the phase of the oscillation.  The effect of
having first squeezing and then oscillation is to have just about any
phase space region evolve into something like the dotted ``cigar'' in
the right panel.  The cigar then undergoes rotation in phase space,
but the entire distribution has a fixed phase of oscillation (up to a
sign).  The degree of phase coherence (or inverse ``cigar thickness'') is
extremely high 
in the real cosmological case because the relevant modes spend a long
time in the squeezing regime. 

\subsection{The results of coherence}
\begin{figure}
\centerline{\psfig{file=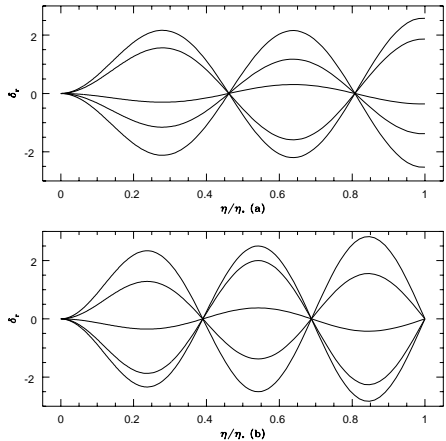,width=2.5in}}
\caption{The evolution of $\delta_\gamma$ for two different wavelengths
(upper and lower panels) as a function of time for an ensemble of
initial conditions.  Each wavelength shows an initial period of growth
(squeezing) followed by oscillations.  The initial squeezing fixes the 
phase of oscillation across the entire ensemble, but different wavelengths 
will have different phases.  The two chosen wavelengths are maximally
out of phase at $\eta_*$ the time of last scattering.}
\label{squeezet}
\end{figure}

Having presented the toy model, let as look at the behavior of the
real cosmological variables.  Figure \ref{squeezet} shows
$\delta_\gamma$ (the photon fluctuations) as a function of (conformal)
time $\eta$ measured 
in units of $\eta_{*}$, the time at last scattering.
Two different
wavelengths are shown (top and bottom panels), and each panel shows
several members of an ensemble of initial conditions.  Each curve
shows an early period of growth (squeezing) followed by
oscillation. The onset of oscillation appears at different times for
the two panels, as each mode ``enters'' $R_J$ at a different time. As
promised, because of the initial squeezing epoch all curves match
onto the oscillatory behavior at the same phase of oscillation (up to a 
sign). Phases can be different for different
wavelengths, as can be seen by comparing the two panels.

\begin{figure}
\centerline{\psfig{file=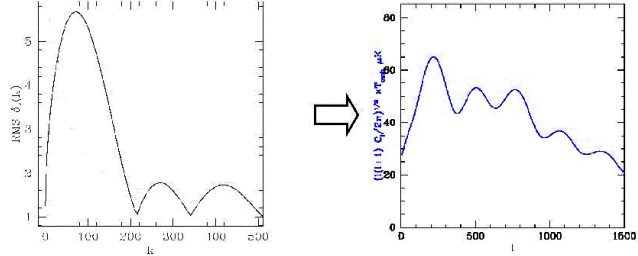,width=3.5in}}
\caption{The left panel shows how temporal phase coherence manifests
itself in the power spectrum for $\delta$ in Fourier space, shown here at one
moment in time. Even after ensemble averaging, some wavelengths are
caught at zeros of their oscillations while others are caught at
maximum amplitudes. These features in the radiation density contrast $\delta$
are the root of the wiggles in the CMB angular power shown in the
right panel.}
\label{pow2cl}
\end{figure}

To a zeroth approximation the event of last scattering simply
releases a snapshot of the photons at that moment and
sends them free-streaming across nearly empty space.  The left panel
of Fig. \ref{pow2cl} (solid curve) shows the mean squared photon
perturbations at the time of last scattering in a standard
inflationary model, vs $k$. 
Note how some wavenumbers have been caught at the nodes of their
oscillations, while others have been caught at maxima.  This feature
is present despite the fact that the curve represents an {\em
ensemble average} because the same phase is locked in for
each member of the ensemble. 

The right panel of Fig. \ref{pow2cl} shows a typical angular power
spectrum of CMB anisotropies produced in an
inflationary scenario. While the right hand plot is not exactly the
same as the left one, it is closely related.  The CMB anisotropy power 
is plotted vs. angular scale instead of Fourier mode, so the
x-axis is ``$l$'' from spherical harmonics rather than $k$. The
transition from $k$ to $l$ space, and the fact that other quantities
besides $\delta_\gamma$ affect the anisotropies both serve to wash out 
the oscillations to some degree (there are no zeros on the right plot, 
for example).  Still the extent to which there {\em are} oscillations
in the CMB power is due to the coherence effects just discussed.

\begin{figure}
\centerline{\psfig{file=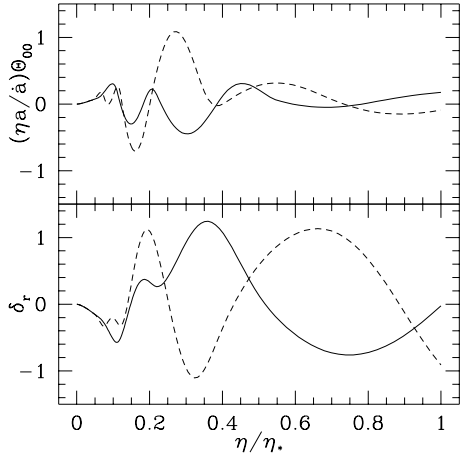,width=2.5in}}
\caption{Decoherence in the defect models: The two curves in the upper 
panel show two different realizations of the effectively random
driving force for a particular mode of $\delta_\gamma$.  The
corresponding evolution of $\delta_\gamma$ is shown in the lower
panel.  There is no phase coherence from one member of the ensemble to 
the next.
}
\label{defnsqz}
\end{figure}

As our understanding of the inflationary predictions has developed,
the defect models of cosmic structure formation have served as a
useful
contrast \cite{VilenkinShellard,AlbrechtCoulsonFerreiraMagueijo}.  In
cosmic defect models there is an added 
matter component (the defects) that behaves in a highly nonlinear way,
starting typically all the way back at the GUT epoch.  This
effectively adds a ``random driving term'' to the equations that is constantly 
driving the other perturbations. These models are called ``active''
models, in contrast to the passive models where all matter evolves in
a linear way at early times. Figure \ref{defnsqz} shows how
despite the clear tendency to oscillate, the phase of oscillation is
randomized by the driving force.  In all known defect models the
randomizing effect wins completely and there are no visible
oscillations in the CMB power.  This comparison will be discussed
further in the next section.

\section{Predictions}
\label{Predictions}

Given a region that has started inflating, and given the inflaton
potential and the coupling of the inflaton to other matter, very
detailed and unambiguous predictions can be made about the initial
conditions of the SBB epoch that follows inflation.  The catch is that at
the moment we do not have a unique candidate for the inflaton and
furthermore we are pretty confused in general about physics at the
relevant energy scales. 

Despite these uncertainties inflation makes a pretty powerful 
set of predictions. I will discuss these predictions here, but first it 
is worth discussing just what one might mean by ``powerful'' predictions. 
From time to time informal discussions of inflation
degenerate into a debate about whether inflation will ever {\em
really} be tested.  The uncertainties about the inflaton mentioned
above are the reason such debates can exist.  As long as there is no
single theory being tested, there is ``wiggle room'' for the theory to 
bend and adapt to pressure from the data.  

Today there are certain
broad expectations of what the inflaton sector might look like, and
despite their breadth, these expectations lead to a set of highly
testable predictions. However, radical departures from these broad
expectations are also possible, and some will argue 
that such departures mean that one is never really testing inflation.

I have little patience for these arguments.  The fundamental currency
of scientific progress is the confirmation or rejections of specific
ideas.  There is no question that the standard picture of cosmic
inflation commits itself to very specific predictions. To 
the extent that any one of these predictions is falsified, a real
revolution will have to occur in our thinking. Furthermore, since most 
of these tests will be realized in the foreseeable future, the testing
of inflation is not a philosophical question.  It is a very real part of
the field of cosmology.

To be absolutely clear on this matter, I have indicated in each topic
below both the predictions of the standard picture and the possible
exotic variations.

\subsection{Flatness}
\begin{figure}
\centerline{\psfig{file=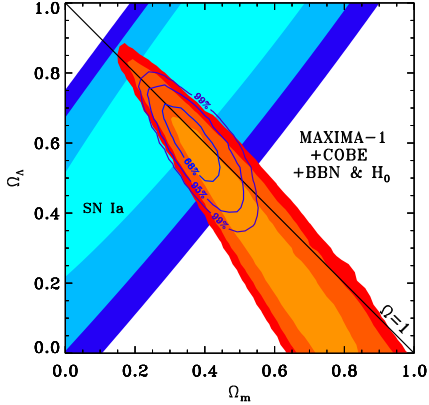,width=2.5in}}
\caption{This is Fig. 4 from [14].  This shows that the
Maxima and Supernova data consistent with a flat Universe (indicated
by the diagonal $\Omega_m + \Omega_{\Lambda} = 1$ line).
}
\label{flat}
\end{figure}

{\bf The Standard Picture:}
In standard inflaton potentials a high degree of adjustment is required to 
achieve a suitably small amplitude for the density perturbations.  It turns 
out that this adjustment also drives the overall time of inflation
far (often exponentially) beyond the ``minimal'' 60 e-foldings.  As a
result, the total curvature of the Universe is precisely zero today in 
inflationary cosmologies, at least as foreseeable observations are
concerned.  Any evidence for a non-zero curvature at the precisions
of realistic experiments will overturn the standard inflationary
picture.  The current data is consistent with a flat Universe as
illustrated in Fig. \ref{flat}.

{\bf Exotic variations:}
The possibility that an open Universe could emerge at the end of 
inflation has been considered for some
time \cite{SteinhardtOpen,LythStewart} For example in
\cite{BucherGoldhaberTurok} a period of ordinary old  
inflation ends by bubble nucleation, and there is subsequently a brief 
period of new inflation inside the bubble.  The degree of openness is 
controlled by the shape of the inflaton potential, which determines
the initial field values inside the bubble and the duration of
inflation afterwards.  The open inflation models suggest an
interesting direction of study if the universe turns out to be open,
but they also represent  a picture that is radically different from
the standard inflationary 
one.  

\subsection{Coherence}

\begin{figure}
\centerline{\psfig{file=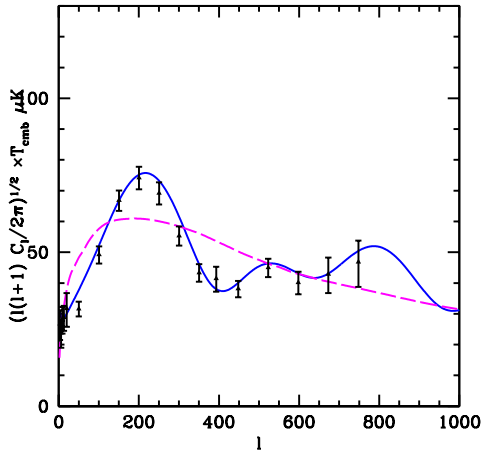,width=2.5in}}
\caption{A current compilation of CMB anisotropy data with curves from
an inflationary model (solid) and an active model (dashed).  The sharp 
peak, a result of the phase coherence in the inflation model is
impossible for realistic active models to produce.
}
\label{defects}
\end{figure}

{\bf The standard picture}
The features in the CMB power spectrum produced by cosmic coherence
(as illustrated in Fig \ref{defects}) are very special signatures of
inflation (and other passive models). Similar features are predicted
in the polarization 
anisotropy power spectrum and in the polarization-temperature
cross-correlations. The absence of such features 
would overturn this standard picture.  In light of this point, the
presence of the so-called ``first Doppler peak'' in current data
(Fig. \ref{defects})  \cite{knox} reflects resounding support for the standard
inflationary picture. Figure \ref{defects} also shows a cosmic defect
based model \cite{WellerBattyeAlbrecht,AlbrechtD2000} that, through
some pretty exotic physics has been manipulated to attempt the
creation of a first Doppler peak. Although the broad shape is
reproduced, the sharp peak and corresponding valley are not.  

{\bf Exotic variations:}
It has been shown that under very extreme conditions, active
perturbations could mimic the features produced by coherence in the
inflationary models \cite{mimic}.  These ``mimic models'' achieved a similar effect
by having the active sources produced impulses in the matter that are
highly coherent and sharply peaked in time.  Despite their efforts to
mimic inflationary perturbations, these models have strikingly
different signatures in the CMB polarization, and thus can be
discriminated from inflationary models using future experiments.
Also, there is no known physical mechanism that could produce the
active sources required for the mimic models to work.

Another possibility is that the perturbations {\em are} produced by
inflation but the oscillatory features in the CMB are hidden due to
features introduced by a wiggly inflaton potential.  It has been shown
 \cite{LewinAlbrecht} that this sort of (extreme) effect would
eventually turn up  
in CMB polarization and temperature-polarization cross-correlation
measurements.  

Another possibility is that inflation solves the flatness, homogeneity 
and horizon
problems, but generates perturbations at an unobservably low
amplitude.  The perturbations then must be generated by some other
(presumably ``active'') mechanism.  Currently there is no good
candidate active mechanism to play this role  \cite{AlbrechtD2000},
but if there was, it would certainly generate clear signatures in the
CMB polarization and temperature-polarization cross-correlation measurements.  

I also note here that a time varying speed of light (VSL) has been proposed
as an alternative to inflation \cite{AlbrechtMagueijo,mof}.  So far all know
realizations of this idea have produced a highly homogeneous universe
which also must be perturbed by some subsequent mechanism.  The building
evidence against active mechanisms of structure formation are also
weakening the case for the VSL idea.

\subsection{Perturbation spectrum and other specifics}

\begin{figure}
\centerline{\psfig{file=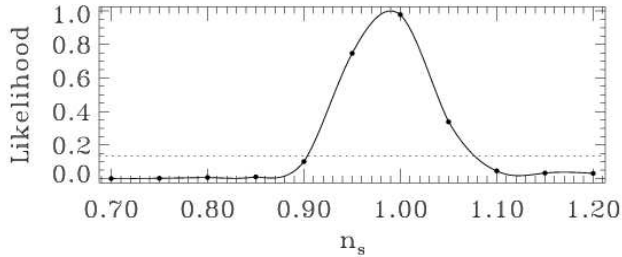,width=3.5in}}
\caption{This is Fig. 2 from [14].  This shows that the
Maxima data are consistent with a scale invariant spectrum
($n=1$).  Other data show a similar level of consistency.
}
\label{tilt}
\end{figure}

One clear prediction from the standard
picture is the ``nearly scale invariant'' spectrum ($\delta_H(k) \approx
{\rm const.}$) discussed in section \ref{Perturbations}.  Strong
deviations (more than several percent) from a scale invariant spectrum
would destroy our standard 
picture.  Fortunately for inflation, there current CMB and other cosmological data gives
substantial support to the idea that the primordial spectrum was indeed 
close to scale invariant. Current data is consistent with a nearly
scale invariant spectrum, as illustrated for example in Fig \ref{tilt}.

Also, the creation of gravitational waves during inflation is a unique
effect \cite{StarGrav}.  While gravity waves from inflation are only
observable with foreseeable experiments for certain inflation
models \cite{ShellardGrav}, we know of no other source of a similar
spectrum of gravity waves.  Thus, if we do observe the right gravity
wave spectrum this would be strong evidence for inflation.

There are a host of specific details of the perturbations that are not 
uniquely specified by all ``Standard Picture'' models.  Some models
have significant contributions form tensor perturbations (gravity
waves) while others do not.  The inflaton potential specifies the
ratio of scalar to tensor amplitudes, but this ratio can be different 
for different models.  

\section{The Big Picture and Open Questions}
\label{tbpaoq}

There are a number of interesting open questions connected with
inflation:

{\bf The origin of the Inflaton:} It is far from clear what the inflaton
actually is and where its potential comes from.  This is intimately
connected with the question of why the perturbations have the amplitude
and spectrum they do.  Currently, there is much confusion about
physics at the relevant energy scales, and thus there is much
speculation about different possible classes of inflaton potentials.
One can hope that a clearer picture will eventually appear as some
deeper theory (such as string/M theory) emerges to dictate the
fundamental laws of physics at the inflation scale.

{\bf Physics of the inflaton:} Having chosen an inflaton
potential, one can calculate the 
perturbations produced during inflation {\em assuming} the relevant
field modes for wavelengths much smaller than the Hubble radius are in
their ground states.  This seems plausible, but it 
would be nice to understand this issue more clearly.  Also, the
inflaton field often takes on values $O(M_P)$.  Will a
yet-to-be-determined theory of quantum gravity introduce large
corrections to our current calculations?

{\bf The cosmological constant problem:} A very important open question is
linked with the cosmological constant 
problem \cite{Carroll,Weinberg,Witten}. The non-zero potential energy of the
inflaton during inflation is very similar to a cosmological constant.
Why the cosmological constant is extremely close to zero today (at
least from a particle physicist's point of view) is perhaps the 
deepest problem in theoretical physics.  One is left wondering whether 
a resolution of this problem could make the cosmological constant and
similar contributions to Einstein's equations identically zero, thus
preventing inflation from every occurring.  Interestingly, current data
is suggesting that there is a non-zero cosmological constant today 
(see for example Fig. \ref{flat} which shows $\Omega_{\Lambda} = 0$ is 
strongly excluded).  Cosmic acceleration today is a very confusing
idea to a theorist, but it actually is helpful to inflation theory in
a number of ways:  Firstly it shows that the laws of physics do allow
a non-zero cosmological constant (or something that behaves in a
similar way).  Also, it is only thanks to the non-zero
$\Omega_{\Lambda}$ that the current data are consistent with a flat
Universe (see Fig. \ref{flat}).

{\bf Wider context and measures:} We 
discussed in Section \ref{tsoi} various ideas about how an
inflating region might emerge from a chaotic start. This is a very
challenging concept to formulate in concrete terms, and not a lot of
progress has been made so far.  In addition, the fact that so many
models are ``eternally inflating'' makes it challenging to define a
unique measure for the tiny fraction of the universe where
inflation actually ends.  It has even been
argued  \cite{LindeMeas1,VilenkinMeas1,LindeMeas2} that these  
measure problems lead to ambiguities in the ultimate predictions from
inflation. 
 
In fact it is quite possible that the ``chaotic'' picture will extend 
to the actual inflaton as
well \cite{AlbrechtMeas,VilenkinMeas2}. Fundamental physics may provide
{\em many} different flat directions that could inflate and
subsequently  reheat, leading to many different versions of the ``Big
Bang'' emerging from a chaotic start.  One then would have to somehow
figure out how to extract concrete predictions out of this apparently
less focused space of possibilities.

I am actually pretty optimistic that in the long run these measure
issues can be resolved \cite{VilenkinMeas3}.  When there are measure
ambiguities it is usually time to look carefully at the actual physics
questions being 
posed... that is, what information we are gathering about the universe 
and how we are gathering it.  It is the observations we actually make 
that ultimately define a measure for our predictions.  Still, we have 
a long way to go before such optimistic comments can be put to the
test.  Currently, it is not even very clear just what space we are
tying to impose a measure on.

I am a strong opponent of the so-called ``anthropic'' arguments.  All of 
science is ultimately a process of addressing conditional probability
questions.  One has the set of all observations and uses a subset of
these to determine one's theory and fix its parameters. Then one can
check if the theoretical predictions match the rest of the
observations. No one really expects we can predict all observations,
without ``using up'' some of them to determine the theory, but the
goal of science is use up as few as possible, thus making as many
predictions as possible.  I believe that this goal must be the only
determining factor in deciding {\em which } observations to sacrifice
to determine the theory, and which to try and predict.  Perhaps the
existence of galaxies will be an effective observation to ``use up'' in
constraining our theories, perhaps it will be the temperature of the
CMB. All that matters in the end is that we predict as much as possible.

Thus far, using ``conditions for life to exist'' has proven an
extremely vague and ineffective tool for pinning down cosmology.  Some
have argued that  ``there must be at least one galaxy'' for life to
exist  \cite{weinberg,tegmark,turok}, but no one really knows what it
takes for life to exist, and certainly if I wanted to try and answer
such a question I would not ask a cosmologist.  Why even mention
life, when one could just as well say ``we know at least one galaxy
exists'' and see what else we can predict?  In many cases (including
 \cite{weinberg,tegmark,turok}) the actual research can be
re-interpreted that way, and my quibble is really only with the
authors' choice of wording. It is these sorts of arguments (carefully
phrased in terms of concrete observations) that could ultimately help
us resolve the measure issues connected with inflation.

\section{Conclusions}

In theoretical cosmology Cosmic Inflation sits at the interface between
the known and unknown.  The end of inflation is designed to connect 
up with the Standard Big Bang model (SBB) which is highly successful at
describing the current observations.  Before inflation lies a
poorly understood chaotic world where apparently anything goes, and
even the laws of physics are not well understood.

Given that inflation is straddling these two worlds, it is not
surprising that many open questions exist.  Despite these open
questions, inflation does a strikingly good job at addressing 
simpler questions that once seemed deep and mysterious when all we had 
was the SBB to work with.  In particular, inflation teaches how 
it is indeed possible for nature to set up the seemingly highly
``unnatural'' initial conditions that are required for the
SBB. Furthermore, inflation gives us a set of predictive signatures which lie 
within scope of realistic tests.  

Already the confrontation with current observations is strongly
favoring inflation.  Upcoming large-scale surveys of the Universe
promise that the pace of these confrontations will pick up over the
next several years. Thus we should know in the foreseeable future
whether we should continue to embrace inflation and build upon its
successes or return to the drawing boards for another try.

\section*{Acknowledgments}

I wish to thank Alex Lewin for comments on this manuscript, and
acknowledge support from DOE grant DE-FG03-91ER40674, and UC
Davis. Special thanks to the Isaac Newton Institute and the organizers
for their hospitality during the this workshop.


\begin{thebibliography}{99}
\bibitem{Guth} A. Guth, {\it Phys. Rev. D} {\bf 23} 347, (1981)
\bibitem{KT} E. Kolb and M. Turner, ``The Early Universe'' Addison
Wesley (1990)
\bibitem{LindeBook} A. Linde, ``Particle Physics and Inflationary
Cosmology'' Harwood, Chur, Switzerland (1990)
\bibitem{Preskill79} J. Preskill, {\it Phys. Rev. Lett.} {\bf 43}
1387, (1979)
\bibitem{GuthWeinberg} A. Guth and E. Weinberg, {\it Nucl. Phys.} {\bf
B212}, 321 (1983)
\bibitem{Linde82} A. Linde, {\it Phys. Lett.} 108B, 389 (1982)
\bibitem{AlbrechtSteinhardt} A. Albrecht and P. Steinhardt, {\it
Phys. Rev. Lett} {\bf 48}, 1220 (1982)
\bibitem{BucherGoldhaberTurok} M. Bucher, A. Goldhaber and N. Turok, {
\it Phys. Rev. D} {\bf 52}, 3314 (1995)
\bibitem{LindeMezhlumian95} A. Linde and A. Mezhlumian, {\it
Phys. Rev. D} {\bf 53} 5538 (1995)
\bibitem{HartleHawking} J. Hartle and S. Hawking {\it Phys. Rev. D}
{\bf 28}, 2960 (1983)
\bibitem{LiddleLyth} {\it Cosmological Inflation and Large-Scale
Structure}  A. Liddle and D. Lyth,  Cambridge, (2000).  See also
D. Lyth and A. Riotto {\it Phys. Rept.} {\bf 314} 1, (2000) and
A. Liddle and D. Lyth, {\it Phys. Rept.} {\bf 231}, 1 (1993) 
\bibitem{VilenkinShellard} A. Vilenkin and P. Shellard ``Cosmic
Strings and other Topological Defects'', Cambridge Univeristy Press,
(1994)
\bibitem{AlbrechtCoulsonFerreiraMagueijo} A. Albrecht, D. Coulson,
P. Ferreira, and J. Magueijo, {\it Phys. Rev. Lett.} {\bf 76}, 1413
(1996)
\bibitem{Maxima} A. Balbi {\it et al} astro-ph/0005124
\bibitem{SteinhardtOpen} P. Steinhardt {\it Nature} {\bf 345}, 41 (1990).
\bibitem{LythStewart} D. Lyth and E Stewart {\it Phys. Lett. } {\bf
B252}, 336 (1990)
\bibitem{knox} {Data compilation provided by L. Knox at}
http://flight.uchicago.edu/knox/radpack.html  
\bibitem{WellerBattyeAlbrecht} J. Weller, R. Battye and A. Albrecht
{\it Phys. Rev D} {\bf 60} 103520, 1999
\bibitem{AlbrechtD2000} A. Albrecht ``The status of cosmic defect
models of structure formation'' Contribution to the XXXVth Rencontres
de Moriond ``Energy Densities in the Universe'' Jan 2000.  In
preparation.
\bibitem{mimic} N. Turok  {\it Phys. Rev. Lett.} {\bf 77}, 4138 (1996) 
\bibitem{LewinAlbrecht} A. Lewin and A. Albrecht astro-ph/9908061
\bibitem{AlbrechtMagueijo}A. Albrecht and J. Magueijo, PRD 59, 43516 (1999)
\bibitem{mof}J. Moffat, International Journal of Physics D, Vol. 2, No. 3
  (1993) 351-365; Foundations of Physics,
  Vol. 23 (1993) 411. 
\bibitem{StarGrav} A. Sarobinsky {\it JETP Lett} {\bf 30}, 682 (1979)
\bibitem{ShellardGrav} R. Battye and P. Shellard {\it
Class. Quant. Grav.} {\bf 13}, A239 (1996)
\bibitem{Carroll} S. Carroll astro-ph/0004075
\bibitem{Weinberg} S. Weinberg astro-ph/0005265 (Talk given at Dark Matter 2000, February, 2000)
\bibitem{Witten}E. Witten, hep-ph/0002297
\bibitem{LindeMeas1}A. Linde, D. Linde and A. Mezhlumian {\it
Phys. Rev. D} {\bf 49}, 1783 (1994)
\bibitem{VilenkinMeas1}V. Vanchurin, A. Vilenkin and S. Winitzki
{\bf Phys. Rev. D} {\bf 61},  083507 (2000)
\bibitem{LindeMeas2} A. Linde, D, Linde, and A.
Mezhlumian {\it Phys. Rev. D} {\bf 54}, 2504 (1996)
\bibitem{AlbrechtMeas} A. Albrecht, In the proceedings of  "The
international workshop on the Birth of the Universe and Fundamental
Forces" Rome, 1994 F. Occionero ed. (Springer Verlag) 1995
\bibitem{VilenkinMeas2} A. Vilenkin {\it Phys. Rev. D} {\bf 52} 3365 (1995)
\bibitem{VilenkinMeas3} See for example A. Vilenkin {\it Phys. Rev. Lett.} {\bf 81} 5501 (1998)
\bibitem{weinberg} S. Weinberg {\it Phys.Rev. D} {\bf 61}, 103505
(2000)
\bibitem{tegmark} M. Tegmark and M. Rees  {\it Astrophys.J.} {\bf 499}, 
 526 (1998)
\bibitem{turok} S. Hawking and N. Turok {\it Phys. Lett.} {\bf B425},
25 (1998) 



\end{thebibliography}
\end{document}